\documentclass[9pt,twocolumn,twoside]{pnas-new}
\usepackage{graphicx,subfigure}
\usepackage[para, online, flushleft]{threeparttable}
\usepackage{wasysym}

\templatetype{pnasresearcharticle} 

\title{Spontaneous emergence of self-replication in chemical reaction systems}

\author[a,1]{Yu Liu}
\author[a]{David Sumpter} 

\affil[a]{Department of Mathematics, Uppsala University, 75105 Uppsala, Sweden}

\leadauthor{Liu} 

\authorcontributions{Author contributions: Y.L. and D.S. designed research; Y.L. performed research; and Y.L. and D.S. wrote the paper.}
\authordeclaration{The authors declare no conflict of interest.}
\correspondingauthor{\textsuperscript{1}To whom correspondence should be addressed. E-mail: yu.liu@math.uu.se}


\begin{abstract}
Explaining the origin of life requires us to explain how self-replication arises. To be specific, how can a self-replicating entity develop spontaneously from a chemical reaction system in which no reaction is self-replicating? Previously proposed mathematical models either supply an explicit framework for a minimal living system or only consider catalyzed reactions, and thus fail to provide a comprehensive theory. We set up a general model for chemical reaction systems that properly accounts for energetics, kinetics and the conservation law. We find that (1) some systems are collectively-catalytic where reactants are transformed into end products with the assistance of intermediates (as in the citric acid cycle), while some others are self-replicating where different parts replicate each other and the system self-replicates as a whole (as in the formose reaction); (2) many alternative chemical universes often contain one or more such systems; (3) it is possible to construct a self-replicating system where the entropy of some parts spontaneously decreases, in a manner similar to that discussed by Schr\"odinger; (4) complex self-replicating molecules can emerge spontaneously and relatively easily from simple chemical reaction systems through a sequence of transitions. Together these results start to explain the origins of prebiotic evolution.
\end{abstract}

\doi{\url{www.pnas.org/cgi/doi/10.1073/pnas.XXXXXXXXXX}}

\begin{document}

\verticaladjustment{-2pt}

\maketitle
\thispagestyle{firststyle}
\ifthenelse{\boolean{shortarticle}}{\ifthenelse{\boolean{singlecolumn}}{\abscontentformatted}{\abscontent}}{}

\dropcap{S}elf-replication is one of the central properties of life \cite{Szathmary06TOO}, and to explain life's origins we need to explain how self-replication arose. It is widely accepted that before DNA, life replicated through RNA molecules \cite{Gilbert86OOL}. However, it remains unclear how the building blocks of RNA, such as nucleotides, became available on the primitive Earth, and even if these building blocks were abundant, it is unclear how they were assembled into the first RNA \cite{Robertson12TOO, Kim11SOC}. It is plausible that self-replication did not originate from a single complex independent self-replicating molecule. In the early stage of evolution, the ``precursor life'' could be very different from what we see today \cite{Robertson12TOO}. For example, in W\"achtersh\"auser's iron-sulfur world hypothesis, the precursor life does not have nucleic acids but consists of a self-replicating (or autocatalytic, in his words) metabolic network \cite{Wachtershauser88BEA}. Another proposal, by Szathm\'ary, is that life evolved from ``holistic limited hereditary replicators'' such as the formose reaction---in which sugar is replicated from formaldehyde---to ``modular unlimited hereditary replicators'' such as RNA and today's DNA \cite{Szathmary99Level, Szathmary95ACO, Maynard95TMT}.

There are lots of biological examples of self-replicating systems that do not rely on a single complex template molecule (as is the case for DNA and RNA molecules). These include: the malic acid cycle (a metabolic path of some bacteria and plants for synthesis of malates); the Calvin cycle in photosynthesis \cite{Ganti03book}; the reductive citric acid cycle for a certain group of chemoautotrophs \cite{Morowitz00TOO, Wachtershauser90EOT}; the metabolic pathways of ATP (as well as some other coenzymes such as NAD$^+$ and CoA) found in many different living organisms \cite{Kun08CIO}; and the whole metabolic reaction network of \textit{E. coli} \cite{Sousa15ASI}. Self-replication has also been identified in non-living systems such as the formose reaction \cite{Szathmary05EPA, Ganti03book} and experiments in labs \cite{Vaidya12SNF, Lincoln09SRO, Ashkenasy04DOA}, including self-replication of nucleotide-based oligomers \cite{Sievers94SOC}.

Several theoreticians have tried to explain the origin of self-replication in terms of a system of coupled chemical reactions \cite{Nghe15PNE}. For example, the chemoton model describes a system composed of three coupled quasi-self-replicating subsystems (metabolism, membrane, and template), which as a whole is able to self-replicate \cite{Ganti03book}. The chemoton can be considered as a model of a minimal living system, but cannot explain how this system spontaneously develops from a soup of simple molecules. The RAF (reflexively autocatalytic and food-generated) theory \cite{Hordijk17CTT, Hordijk04DAS, Steel00TEO}, extended from Kauffman's autocatalytic sets theory \cite{Kauffman86ASO}, is also a step in the direction of understanding origins of self-replication. A set of chemical reactions is RAF if (1) every reaction in this set is catalyzed by at least one molecule involved in this set, and (2) every molecule involved in this set can be produced from a small food molecule set. RAF sets are shown to be able to readily emerge from a set of chemical reactions, and always consist of smaller RAF sets, demonstrating the capability to evolve \cite{Hordijk13ASF, Hordijk12TSO}. Other similar models also contributed to this theory \cite{Hordijk16EOA, Vasas12EBG, Jain01AMF, Wills00SAI}, and many of the biological observations mentioned in the previous paragraph can be put into the framework of RAF theory \cite{Sousa15ASI, Vaidya12SNF, Lincoln09SRO, Ashkenasy04DOA, Sievers94SOC}.

Although RAF theory is interesting, it has limitations as an explanation of self-replication. Firstly, the theory stipulates that every reaction in the set is catalyzed. Even though catalyzed reactions are very common in living systems, not every reaction involves a catalyst (e.g., condensation reactions \cite{Szathmary95ACO, Fontana94TAO}). In the early stages of biological evolution, probably no reaction required sophisticated biotic catalysts (e.g., enzymes) \cite{Sousa15ASI, Morowitz00TOO}, so there is a strong motivation not to include these enzymes \cite{Rasmussen16GML, Levy03EGB, Sievers94SOC} or even catalysts as given in a model of the origins of self-replication. Uncatalyzed reactions have significant effects on the dynamics of the whole system, e.g., ``innovating'' new species of molecules and then triggering other RAF sets \cite{Hordijk16ASI, Farmer86ARO}. The second concern with RAF theory is that it is a purely graph-theoretic approach. As a result, there is no constraint on how chemical reactions are constructed and coupled (although it gives the theory much freedom, the construction of reaction systems is too arbitrary to investigate the model systematically). Extra assumptions need to be made about chemical kinetics in order to investigate the dynamics of how populations of molecules change over time \cite{Smith14ASI}. Here we see another reason to relax the assumption of studying only catalyzed reaction: catalyzed reactions are never elementary reactions, so the kinetics cannot be simply calculated from the reaction equations \cite{Atkinstextbook}.

Any theoretical approach to the origin of self-replication should explicitly include energetics \cite{Sousa15ASI, Martin14EAL}, an aspect which is missing from all these models and theories above. There are several reasons for this. Firstly, energetics (e.g., Gibbs energy) determines whether a chemical reaction is spontaneous, so to investigate the spontaneity of the emergence of life, it has to be considered. Secondly, in order to concretely discuss the issue---famously put forward by Schr\"odinger---that life maintains its order by feeding on ``negative entropy'' \cite{Schrodinger44WIL}, energetics has to be explicitly taken into account: entropy is negatively related to the thermodynamic free energy (which is Gibbs energy in the scenario of constant pressure and temperature). It should be noted that the relationship between life and entropy is investigated in different ways and contexts \cite{Amend13TEO, Branscomb13TAB, England13SPO}. For example, Branscomb explained the specific mechanisms to increase thermodynamic free energy in two real-world biochemical scenarios \cite{Branscomb13TAB}: the hypothesized ``alkaline hydrothermal vent'' \cite{Russell93OTE} on the prebiotic Earth, and the system where the Ferredoxin I protein translocates protons. In the context of statistical physics, a lower limit was derived for the amount of heat generated in a non-equilibrium system where a process of self-replication occurs \cite{England13SPO}.

In this paper, we set up a general model for chemical reaction systems that properly accounts for energetics, kinetics and the conservation law. Catalysts are not explicitly included in the model, but we later find that catalysis can emerge, along with self-replication and potentially complex molecules, in our system.

\section*{Model}

\subsection*{Chemical reaction system}

We model a well-mixed soup of molecules, each of which is defined by its integer mass, $i$. A molecule's type is thus denoted $\overline{i}$. Only two types of reaction are possible: synthesis of two molecules to create a molecule of greater mass (e.g., $\overline{2} + \overline{4} \rightarrow \overline{6}$) and decomposition into two molecules to create two molecules of lower mass (e.g., $\overline{6} \rightarrow \overline{1} + \overline{5}$). All reactions that conserve mass---the total mass on the left-hand side of the equation adds up to those on the right-hand side---can occur. For convenience, we define a reaction pair to be a reaction and its corresponding reverse reaction.

\begin{figure}[htbp] \centering
  \subfigure{ \label{fig:ddag1}
    \includegraphics[width = 0.47\linewidth]{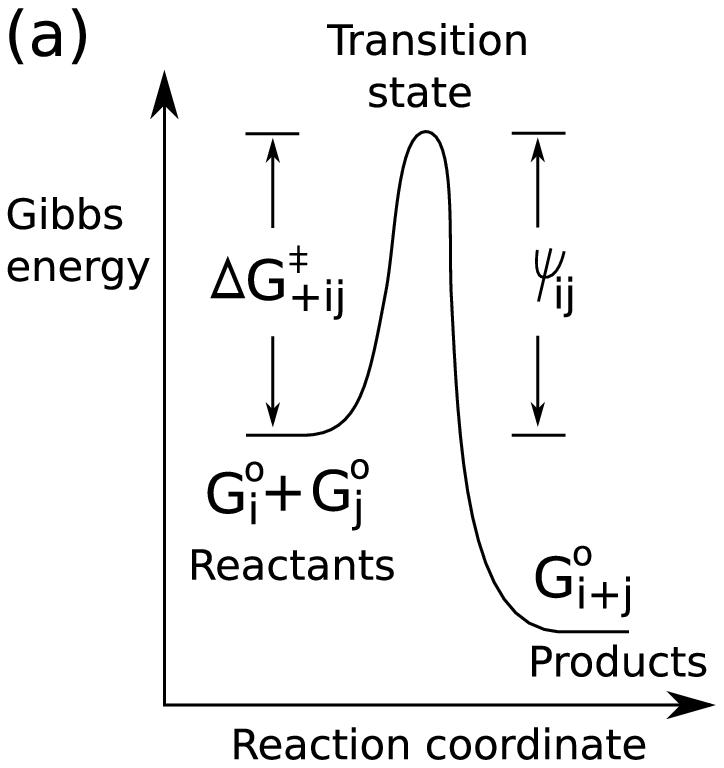}}
  \subfigure{ \label{fig:ddag2}
    \includegraphics[width = 0.47\linewidth]{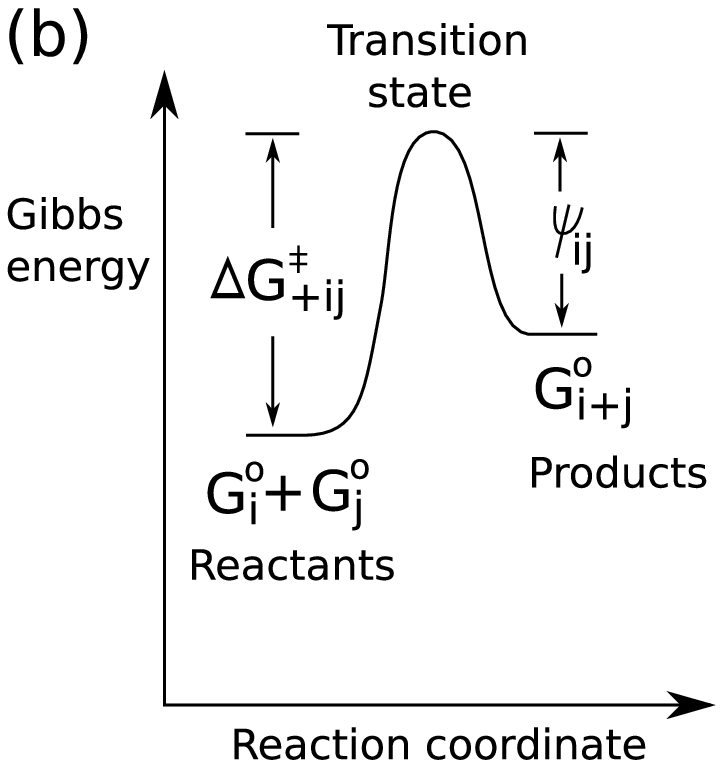}}
    \caption{Diagram of Gibbs energy for a synthesis reaction $\overline{i} + \overline{j} \rightarrow \overline{i+j}$. (a) For the case that the synthesis reaction is spontaneous, i.e., $G^\circ_i + G^\circ_j > G^\circ_{i+j}$. (b) For the case that it is non-spontaneous, i.e., $G^\circ_i + G^\circ_j \leqslant G^\circ_{i+j}$.}
\label{fig:ddag}
\end{figure}

Each type of molecule $\overline{i}$ has its own standard Gibbs energy of formation $G^\circ_i$. Then, as illustrated in Fig. \ref{fig:ddag},
a reaction is either spontaneous---meaning that the total standard Gibbs energy of formation of the reactants is greater than that of the products---i.e., $G^\circ_i + G^\circ_j > G^\circ_{i+j}$, or non-spontaneous, i.e., $G^\circ_i + G^\circ_j \leqslant G^\circ_{i+j}$. If one reaction in a reaction pair is spontaneous, the other is non-spontaneous, and vice versa. 

According to transition state theory \cite{Atkinstextbook}, the reactants have to overcome the Gibbs energy of activation (namely $\Delta G^\ddag_{+ij}$ in Fig. \ref{fig:ddag}) in order for the reaction to occur. In the model, any reaction pair is either \textit{low-barrier}---the energy barrier (corresponding to $\psi_{ij}$ in Fig. \ref{fig:ddag}) is low and the reaction rate is thus high---or \textit{high-barrier}. In our system all reactions are possible, but to define a specific chemical reaction system we write a list of low-barrier (and spontaneous) reactions only. For example,
\begin{equation} \label{eq:Krebs}
\begin{cases} \begin{aligned}
 \overline{5}       & \rightarrow \overline{1} + \overline{4} \\
 \overline{6}       & \rightarrow \overline{1} + \overline{5} \\
 \overline{2} + \overline{4}       & \rightarrow \overline{6}
\end{aligned} \end{cases} \end{equation}
is a model of the citric acid cycle in cellular respiration, simplified so that only carbon-changing reactions are included \cite{Ganti03book}. Specifically, molecule $\overline{1}$ stands for carbon dioxide, $\overline{2}$ for acetyl-CoA, $\overline{4}$ for oxaloacetic acid, $\overline{5}$ for $\alpha$-Ketoglutaric acid, and $\overline{6}$ for citric acid, respectively.

It is also possible to model the formose reaction \cite{Ganti03book} which involves the formation of sugars from formaldehyde,
\begin{equation} \label{eq:formose}
\begin{cases} \begin{aligned}
 \overline{1} + \overline{2}       & \rightarrow \overline{3} \\
  \overline{1} + \overline{3}       & \rightarrow \overline{4} \\
 \overline{4}       & \rightarrow \overline{2} + \overline{2} 
\end{aligned} \end{cases} \end{equation}
Again, only carbon-changing reactions are considered. Specifically, molecule $\overline{1}$ stands for formaldehyde, $\overline{2}$ for glycolaldehyde, $\overline{3}$ for glyceraldehyde, and $\overline{4}$ for tetrose, respectively. These two reaction systems will serve as special cases in our study, but we will cover a wide range of different systems.

Note that some chemical reaction systems are not physically possible, e.g.,
\begin{equation*} \begin{cases} \begin{aligned}
 \overline{2}       & \rightarrow \overline{1} + \overline{1} \\
 \overline{1} + \overline{2}       & \rightarrow \overline{3} \\
 \overline{1} + \overline{3}       & \rightarrow \overline{4} \\
 \overline{4}       & \rightarrow \overline{2} + \overline{2} 
\end{aligned} \end{cases} \end{equation*}
where by adding up these low-barrier reactions, all molecules are cancelled out. As a consequence, one cannot find proper Gibbs energy for each molecule such that each reaction above is spontaneous. We are only interested in physically possible systems.

\subsection*{Kinetics}

We now specify a general model for the kinetics of our chemical system, under the following assumptions: (1) All molecules are ideally gaseous; (2) the whole system is kept at constant pressure and temperature; (3) every possible reaction is elementary. The derivation follows the rate law and transition state theory \cite{Atkinstextbook}. Here we cover the key points, and a full derivation is given in Supplementary Information (SI) section S1.

For any synthesis reaction $\overline{i} + \overline{j} \rightarrow \overline{i+j}$, the reaction rate in unit $s^{-1}$ is
\begin{equation*} \label{eq:synrate}
  \gamma_{+ij} = \beta \exp(-\kappa \Delta G^\ddag_{+ij} ) \cdot N_i \cdot N_j / ( S + N ) 
\end{equation*}
where the subscript $_{+ij}$ stands for the synthesis reaction, $\beta$ and $\kappa$ are constants, $N_i$ is the number of molecules $\overline{i}$ in the system, $N_j$ is the number of $\overline{j}$, $S$ is the number of solvent molecules, determining the global rate at which molecules interact, $N$ is the number of all the molecules except for solvent molecules, and $\Delta G^\ddag_{+ij}$, as shown in Fig. \ref{fig:ddag}, is defined as
\begin{equation*}
  \Delta G^\ddag_{+ij} = \begin{cases}
    \psi_{ij}, & \text{if~} G^\circ_i + G^\circ_j > G^\circ_{i+j} \\
    \psi_{ij} + G^\circ_{i+j} - (G^\circ_i + G^\circ_j), & \text{if~} G^\circ_i + G^\circ_j \leqslant G^\circ_{i+j}
\end{cases} \end{equation*}
When implementing the model, we set values of $\psi_{ij}$ (positive and finite) for each reaction pair and $G^\circ_i$ for each molecule. Together, these give a unique value for $\Delta G^\ddag_{+ij}$.

Likewise, for any decomposition reaction $\overline{i+j} \rightarrow \overline{i} + \overline{j}$, the reaction rate in unit $s^{-1}$ is
\begin{equation*} \label{eq:decrate}
  \gamma_{-ij} = \beta \exp{(-\kappa \Delta G^\ddag_{-ij})} \cdot N_{i+j}
\end{equation*}
where the subscript $_{-ij}$ stands for this decomposition reaction, and
\begin{equation*}
  \Delta G^\ddag_{-ij} = \begin{cases}
    \psi_{ij}, & \text{if~} G^\circ_{i+j} > G^\circ_i + G^\circ_j, \\
    \psi_{ij} + (G^\circ_i + G^\circ_j) - G^\circ_{i+j}, & \text{if~} G^\circ_{i+j} \leqslant G^\circ_i + G^\circ_j.
\end{cases} \end{equation*}

Due to the fact that the transition state of a reaction pair $\overline{i} + \overline{j} \rightarrow \overline{i+j}$ and $\overline{i+j} \rightarrow \overline{i} + \overline{j}$ is identical, by setting $\psi_{ij}$ both $\Delta G^\ddag_{+ij}$ and $\Delta G^\ddag_{-ij}$ are uniquely determined. In addition, by setting $\psi_{ij}$ large or small, we can easily make the reaction pair low-barrier or high-barrier.

\subsection*{Simulation of dynamics}

We take the citric acid cycle \eqref{eq:Krebs} as an example to illustrate how to set up the simulation experiment.

(1) Set up all the constants. Assume that the constant pressure the system is kept at is $100~kPa$ and the constant temperature is $298.15~K$, so we obtain that $\beta \approx 6.21\times10^{12}~s^{-1}$ and $\kappa \approx 0.403~mol/kJ$.

(2) Set $G^\circ_i$ for each type of molecule (up to $\overline{6}$ in this case), to make sure that the three reactions are spontaneous, that is, $G^\circ_5 > G^\circ_1 + G^\circ_4$, $G^\circ_6 > G^\circ_1 + G^\circ_5 $ and $G^\circ_2 + G^\circ_4 > G^\circ_6$. Here we set $G^\circ_1 = -780$, $G^\circ_2 = -500$, $G^\circ_3 = -490$, $G^\circ_4 = -190$, $G^\circ_5 = -830$ and $G^\circ_6 = -900$, all in unit $kJ/mol$. These values are set in the range of normal chemical substances' Gibbs energy of formation \cite{Atkinstextbook}, roughly in the range $(-1500, 300)$. Note that the choice is not unique and a wide range of choices can be made to allow the system to work.

(3) Set $\psi_{14} = \psi_{15} = \psi_{24} = 10$ for these three low-barrier reactions, and $\psi_{ij}=100$ for all other reaction pairs (all in units $kJ/mol$).

(4) Assume that there is an unlimited reservoir of resource molecule $\overline{2}$. That is, whenever a molecule $\overline{2}$ is consumed or produced, it is replenished or removed so that the number of $\overline{2}$ always keeps as a constant $Q = 1000$. This setting makes biological sense if we consider a system separated from the unlimited reservoir by a ``wall''. As long as some resource molecules are consumed, more will enter the system driven by the chemical gradient.

(5) We denote the number of molecule $\overline{i}$ at time $t$ as $N_i(t)$. Initially, $N_4(0) = 1$, and $N_1(0) = N_3(0) = N_5(0) = N_6(0) = 0$. Molecule $\overline{4}$ triggers the chemical reaction system. We also set $S = 1\times10^6$, which is much larger than $Q$, so that we are in the dilute limit.

We then use the standard Gillespie Algorithm to simulate the dynamics (see details in SI section S2). In addition, we also construct ordinary differential equations (ODEs) to describe the mean-field dynamics (see details in SI section S3).

\section*{Results}

\subsection*{Collectively-catalytic system}

We first simulate the citric acid cycle \eqref{eq:Krebs} and observe that $N_1(t)$ increases linearly. Molecules $\overline{4}, \overline{5}$ and $\overline{6}$ are involved in a cycle of reactions, but the total number is constant. The SI section S4 shows this dynamics and ODEs solutions in more detail. We also observe that each reaction in \eqref{eq:Krebs} occurs approximately the same number of times. We can add up these reactions, cancel out the molecules appeared on both sides of the reaction (in this case, molecule $\overline{4}$, $\overline{5}$ and $\overline{6}$), and then obtain the overall reaction of the system $\overline{2} \rightarrow \overline{1} + \overline{1}$. In itself, the reaction $\overline{2} \rightarrow \overline{1} + \overline{1}$ is high-barrier, so its reaction rate is extremely low, but through the whole system the actual rate of the overall reaction is several billion times larger (SI section S4). We call system \eqref{eq:Krebs} \textit{collectively-catalytic system}, since the overall reaction $\overline{2} \rightarrow \overline{1} + \overline{1}$ is catalyzed by molecule $\overline{4}$, $\overline{5}$ and $\overline{6}$. Note that this outcome is consistent with the biological observation that the citric acid cycle consumes acetyl-CoA and produces carbon dioxide as a waste product.

The linear growth of molecules $\overline{1}$ is because: (1) $\overline{1}$ is an end product which cannot be used by these low-barrier reactions; (2) the number of resource molecules $\overline{2}$ is constant; and (3) no additional molecule $\overline{4}$, $\overline{5}$ and $\overline{6}$ can be produced through these low-barrier reactions (by noting that the number of $\overline{4}$, $\overline{5}$ and $\overline{6}$ on the right-hand side of the reaction system \eqref{eq:Krebs} is the same as the number on the left-hand side). We can use these observations to give a rigorous set of criteria for collectively-catalytic systems. We start by defining an \textit{intermediate molecule} to be any molecule that appears on both the reactant side and the product side. In general, the following stoichiometric criteria are sufficient (but not necessary) to show that a physically possible chemical reaction system, given supplies of resource molecules, is collectively-catalytic: (1) For every low-barrier reaction, at least one type of its reactants comes from the products of other low-barrier reactions (called the criterion for self-driven); (2) by adding up all the low-barrier reactions, for every type of intermediate molecule, the number of times it appears on the reactant side and on the product side is the same (called the criterion for balanced-cancelling).

The citric acid cycle \eqref{eq:Krebs} thus satisfies all these criteria, but we also find that other systems fulfill these criteria too (in fact, any single catalytic reaction can be written as a collectively-catalytic system: see SI section S5 for details). The criteria above give us a way of discerning whether or not a system is collectively-catalytic based on stoichiometry alone, without the need to investigate its dynamics.

\subsection*{Self-replicating system}

\begin{figure}[htbp] \centering
    \includegraphics[width = 0.9\linewidth]{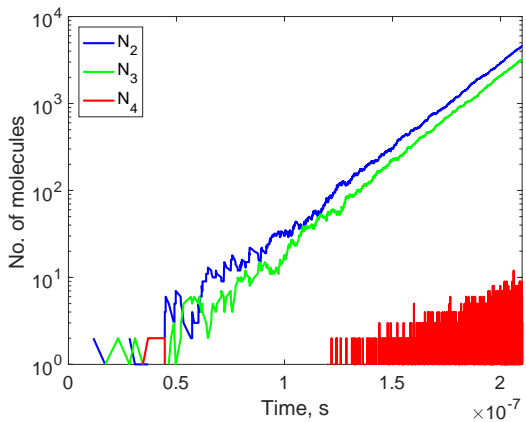}
\caption{Dynamics of the formose reaction \eqref{eq:formose} in log-normal scale, i.e., x-axis is in normal scale and y-axis is in logarithmic scale. It is not so clear that $N_4(t)$ grows exponentially because $N_4(t)$ is always small. But in solutions of ODEs, we see it clearly (SI Fig. S3a). Note that $N_4(t)$ fluctuates frequently between $0$ and small numbers, so the curve looks like a block. We set $G^\circ_1 = 220$, $G^\circ_2 = -760$, $G^\circ_3 = -970$, $G^\circ_4 = -1160$, $N_1(t) = Q$, $N_2(0) = 1$ and $N_3(0) = N_4(0) = 0$.}
\label{fig:dynformose}
\end{figure}

We now look at the dynamics of the formose reaction \eqref{eq:formose} given resource molecule $\overline{1}$. The result of a simulation is shown in Fig. \ref{fig:dynformose}.
All three intermediate molecules ($N_2(t)$, $N_3(t)$ and $N_4(t)$) increase exponentially. The solutions of the corresponding ODEs are consistent with the simulations (SI section S6).

The fact that $\overline{2}$ grows exponentially can be seen directly by adding up the three low-barrier reactions to obtain $\overline{1} + \overline{1} + \overline{2} \rightarrow \overline{2} + \overline{2}$. It indicates that if one $\overline{2}$ is present beforehand, one extra $\overline{2}$ can be produced, by transforming two of $\overline{1}$. Then, the additional $\overline{2}$ is further used by the system, and more $\overline{2}$s are produced. Although the molecules $\overline{3}$ and $\overline{4}$ are canceled out when we add up the reactions, they also grow exponentially (with $\overline{4}$ increasing much slower than the other two). The reason is that the actual reaction rate of each low-barrier reaction is not the same (see SI section S6). This observation is important since it illustrates that it is not just one type of molecule that grows exponentially in such systems, but all the intermediate molecules.

We define a \textit{self-replicating system}, of which \eqref{eq:formose} is an example, to be a system in which at least one type of molecule is replicated. By investigating various self-replicating systems, we find that not only exponential but also super-exponential growth is observed (see examples in SI section S7). The dynamics indicates that the reactions in self-replicating systems become faster and faster. This is a very special property compared to the collectively-catalytic system, where the overall reaction rate keeps constant, e.g., the citric acid cycle \eqref{eq:Krebs}.

In general, the following stoichiometric criteria are sufficient (but not necessary) to show that a physically possible chemical reaction system, given supplies of resource molecules, is self-replicating: (1) The criterion for self-driven (mentioned in last section) is satisfied; (2) there are some types of intermediate molecules, and the number of times it appears on the reactant side is less than that on the product side (called the criterion for overproduction); (3) there is no type of intermediate molecules that the number of times it appears on the reactant side is larger than that on the product side (called the criterion for no-overintake). The formose reaction \eqref{eq:formose} satisfies all these criteria.

\subsection*{Collectively-catalytic and self-replicating systems are common}

A natural question to ask is how common these reaction systems are, such as the citric acid cycle and the formose reaction. Specifically, if we construct alternative chemical universes where arbitrarily choose which reaction is low-barrier, how many of the resulting chemical universes contain collectively-catalytic or self-replicating systems? To answer this question we start by using the criterion shared by both of them, that they are self-driven. Table \ref{tab:NumCCSR} shows the numbers for different $L$, which is set to be the mass of the largest molecule in the chemical universe in question, in order to have a measurement of the number of all chemical reaction systems.
\begin{table}[htbp] \centering
  \caption{Number of physically possible universes that contain self-driven, collectively-catalytic or self-replicating systems}
\begin{threeparttable}
\begin{tabular}{l |  r r r r}
        & No. physically                & No. uni. cont.           & Lower bound            & Lower bound \\
 $L$ & possible uni. \tnote{1}   & self-driven \tnote{2}  & uni. cont. CC \tnote{3}  & uni. cont. SR \tnote{4}  \\\hline
    $4$ & $79$          & $8~(10\%)$            & $0~(~~~0\permil)$      & $2~(25.3\permil)$ \\ 
    $5$ & $681$        & $152~(22\%)$        & $5~(7.3\permil)$      & $10~(14.7\permil)$ \\ 
    $6$ & $16,825$   & $6,886~(41\%)$     & $21~(1.2\permil)$    & $74~(~~4.4\permil)$ \\ 
    $7$ & $401,445$ & $232,552~(58\%)$ & $184~(0.5\permil)$  & $642~(~~1.6{\permil})$
\end{tabular}
\begin{tablenotes}
    \footnotesize
    \item[1] Number of physically possible universes. $^2$ Number of physically possible universes that contain self-driven systems. $^3$ Lower bound on the number of physically possible universes that contain collectively-catalytic systems. $^4$ Lower bound on the number of physically possible universes that contain self-replicating systems. All the percentages are calculated with respect to the number of all physically possible universes.
\end{tablenotes}
\end{threeparttable}
\label{tab:NumCCSR}
\end{table}
For example, when $L = 6$, there are in total $\overline{1}$, $\overline{2}$, $\cdots$, $\overline{6}$ six types of molecules, and thus $9$ reaction pairs. By choosing which reaction is low-barrier, we can construct $\sum_{l=0}^9 (^9_l) \cdot 2^l = 19683$ alternative chemical universes, $16825$ of which turn out to be physically possible. Using the criterion for self-driven given above, we find that $6886~(41\%)$ of all the physically possible universes contain self-driven systems. This percentage increases with $L$, which indicates that self-driven systems are common, and more common in systems involving more types of molecules.

However, we cannot be sure that these self-driven systems are collectively-catalytic or self-replicating, and there is a third type of self-driven system that is non-sustaining system (see SI section S8 for details of non-sustaining system and section S9 for more details of classification for self-driven system). Nevertheless, we can use the stoichiometric criteria mentioned above (note that these criteria are sufficient but not necessary) to give a lower bound on the number of chemical universes that contain collectively-catalytic or self-replicating systems. That is, for a self-driven system, a system is collectively-catalytic if it satisfies the criterion for balanced-canceling, or self-replicating if it satisfies both the criteria for overproduction and no-overintake. 

Using the stoichiometric criteria we find that the number of chemical universes containing collectively-catalytic or self-replicating systems increases with $L$, although the percentage decreases. This is a lower bound, so does not mean that the actual number of chemical universes containing collectively-catalytic or self-replicating systems decreases. Establishing a firm relationship between the number of chemicals ($L$) and self-replication will involve simulating dynamics of all systems.

\subsection*{How can life maintain low entropy?}

According to the second law of thermodynamics, the total entropy of an isolated system never spontaneously decreases over time. Life, thought of as an open system as opposed to an isolated one, is able to maintain order, i.e., maintain a relatively low entropy level. Schr\"odinger suggested that this is achieved by life ``feeding on negative entropy'' \cite{Schrodinger44WIL}. His question is how this can happen spontaneously. But before we answer this we first need a way to discuss this question concretely and more quantitively.

Under the framework of our model, if we simply consider life as some self-replicating entity, we should then ask: Is it possible that a self-replicating system spontaneously increases its Gibbs energy or at least keeps it unchanged (we first note that in the scenario of constant pressure and temperature, the decrease of entropy corresponds to the increase of Gibbs energy as $G = H - TS$ where $G$ is Gibbs energy, $H$ is enthalpy, $T$ is temperature and $S$ is entropy)? Let us consider the self-replicating system \eqref{eq:entropy2213}, given the resource molecule $\overline{2}$:
\begin{equation} \label{eq:entropy2213}
\begin{cases} \begin{aligned}
 \overline{5}       & \rightarrow \overline{1} + \overline{4} \\
 \overline{2} + \overline{3}       & \rightarrow \overline{5} \\
 \overline{2} + \overline{4}       & \rightarrow \overline{6} \\
 \overline{6}       & \rightarrow \overline{3} + \overline{3}
\end{aligned} \end{cases} \end{equation}
The simulation shows that $N_1(t)$, $N_3(t)$ and $N_4(t)$ increase exponentially, as well as $N_5(t)$ and $N_6(t)$ (although they are very small, see SI section S10). Molecule $\overline{1}$ is the end product, while $\overline{3}$, $\overline{4}$, $\overline{5}$ and $\overline{6}$ are replicated through the whole system. We then consider that the ``living'' system consists of the self-replicating part (namely all of molecules $\overline{3}$, $\overline{4}$, $\overline{5}$ and $\overline{6}$) and the resource molecules in the system.

\begin{figure}[h] \centering
    \includegraphics[width = 0.9\linewidth]{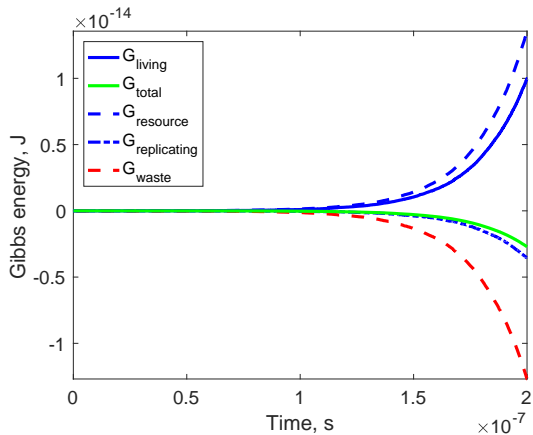}
\caption{Time series of Gibbs energy of the self-replicating system \eqref{eq:entropy2213}. We set $G^\circ_1 = -800$, $G^\circ_2 = -500$, $G^\circ_3 = -400$, $G^\circ_4 = -200$, $G^\circ_5 = -950$, $G^\circ_6 = -750$, $N_2(t) = Q$, $N_3(0) = 1$ and initially other molecules none.}
\label{fig:Gibbs}
\end{figure}

Now we investigate how Gibbs energy of the system changes. We set Gibbs energy of the initial system to zero (as the reference point), since only relative quantity matters. Therefore, Gibbs energy of the self-replicating part is $G_{replicating}(t) = \sum_{i=3}^{6} N_i(t) \cdot G_i^\circ$. Gibbs energy of the resource molecules in the system is $G_{resource}(t) = -F_2(t) \cdot G_2^\circ$ where $F_2(t)$ is the number of resource molecule $\overline{2}$ ever consumed till time $t$. Gibbs energy of the waste is $G_{waste}(t) = N_1(t) \cdot G_1^\circ$. Then, Gibbs energy of the living system is $G_{living}(t) = G_{replicating}(t) + G_{resource}(t)$. As shown in Fig. \ref{fig:Gibbs}, $G_{living}(t)$ increases while $G_{total}(t) = G_{living}(t) + G_{waste}(t)$ decreases.

We have thus given an explicit example of a self-replicating ``living'' system that spontaneously consumes the resources to increase its own Gibbs energy. Note that our system, as defined, is a well-mixed gas system. So the waste molecules $\overline{1}$ are not automatically separated from other molecules, and Gibbs energy of the gas-mixing process is neglected in the calculation above. However, the contribution of the fact of well-mixed gas is relatively small (see details in SI section S10). So it is still possible for a self-replicating system to spontaneously increase Gibbs energy or at least keep it unchanged.

\subsection*{Spontaneous evolution from simple towards complex}

Nature provides many examples of the spontaneous evolution from simple towards complex. Is it possible to construct a system showing the similar process? Imagine there is a chemical reaction system composed of the following low-barrier reactions, given an infinite reservoir of only resource molecules $\overline{1}$.
\begin{table}[h] \centering
\begin{tabular}{ll} \hline
\parbox{3.8cm}{ ~\\
A self-replicating system \\
(i.e., formose reaction \eqref{eq:formose}):
\begin{equation*}
\begin{cases} \begin{aligned}
 \overline{1} + \overline{2}       & \rightarrow \overline{3} \\
  \overline{1} + \overline{3}       & \rightarrow \overline{4} \\
 \overline{4}       & \rightarrow \overline{2} + \overline{2} 
\end{aligned} \end{cases} \end{equation*}

A collectively-catalytic system:
\begin{equation} \label{eq:sys336}
\begin{cases} \begin{aligned}
 \overline{3} + \overline{5}       & \rightarrow \overline{8} \\
 \overline{3} + \overline{8}       & \rightarrow \overline{11} \\
 \overline{11}       & \rightarrow \overline{5} + \overline{6}
\end{aligned} \end{cases} \end{equation} }  &
 
\parbox{3.8cm}{ 
A self-replicating system:
\begin{equation} \label{eq:sequence1}
 \begin{cases} \begin{aligned}
 \overline{1} + \overline{12}       & \rightarrow \overline{13} \\
 \overline{1} + \overline{13}       & \rightarrow \overline{14} \\
 \cdots \\
 \overline{1} + \overline{i}       & \rightarrow \overline{i+1} \\
 (i = 14, & ~15, \cdots, 22) \\
 \cdots \\
 \overline{1} + \overline{23}       & \rightarrow \overline{24} \\
 \overline{24}       & \rightarrow \overline{12} + \overline{12}
\end{aligned} \end{cases} \end{equation} } \\ \hline
\end{tabular}
\end{table}
The first three reactions constitute the formose reaction \eqref{eq:formose}, given the resource molecule $\overline{1}$. The three reactions in \eqref{eq:sys336} constitute a collectively-catalytic system, given the resource $\overline{3}$. The thirteen reactions in \eqref{eq:sequence1} constitute a self-replicating system, given the resource $\overline{1}$. Is it possible that lots of complex molecules, such as $\overline{12}$, $\overline{13}$ and $\overline{14}$, are produced in the end?

The answer in this case is yes. If the first $\overline{12}$ is produced, the self-replicating system \eqref{eq:sequence1} will be triggered, and consequently $N_{12}$ will grow exponentially, as well as $N_{13}$, $N_{14}$, $\cdots$, $N_{23}$. But how is the first $\overline{12}$ produced? There are three stages: (1) Initially when there are only lots of $\overline{1}$ but nothing else, the system stays almost unchanged for a very long time since no low-barrier reaction could occur. Occasionally, by the high-barrier reaction $\overline{1} + \overline{1} \rightarrow \overline{2}$, one molecule $\overline{2}$ is produced. The self-replicating system \eqref{eq:formose} is triggered, and then $N_2$ and $N_3$ grow exponentially. Very quickly there are lots of $\overline{2}$ and $\overline{3}$. (2) After a relatively long ``boring'' period, the first $\overline{5}$ is produced by the high-barrier reaction $\overline{2} + \overline{3} \rightarrow \overline{5}$. Then the collectively-catalytic system \eqref{eq:sys336} is triggered, and $N_6$ grows. Very soon there are lots of $\overline{6}$. (3) After that, occasionally one $\overline{12}$ is produced by the high-barrier reaction $\overline{6} + \overline{6} \rightarrow \overline{12}$.

One might naively believe that the first $\overline{12}$ can be produced by other reactions without the need for self-replicating and collectively-catalytic systems. This is not the case. Despite an abundance of molecules $\overline{1}$ which could be used to ``assemble'' an initial $\overline{12}$, the production of the first $\overline{12}$ requires high-barrier reactions, e.g., $\overline{6} + \overline{6} \rightarrow \overline{12}$. It is only when both the other self-replicating and collective-catalytic systems are producing $\overline{6}$ in sufficient numbers that one such reaction is sufficiently likely to occur. At this point a $\overline{6} + \overline{6} \rightarrow \overline{12}$ can occur, despite it being a high-barrier reaction, because of the abundance of $\overline{6}$ molecules. 

The previous stage-by-stage procedure could be a general model of how chemical reaction systems evolve towards complex: a relatively simple innovation triggers some self-replicating or collectively-catalytic systems and then a large number of new types of molecules are produced, paving the way for other innovations.

Meanwhile, the more types of molecules, the more probable to have reactions which are high-barrier before becoming practically low-barrier. For example, in the formose reaction \eqref{eq:formose}, if the reaction $\overline{1} + \overline{3} \rightarrow \overline{4}$ is high-barrier, the system is not self-replicating. But imagine that the following three reactions are low-barrier,
\begin{equation*} \begin{cases} \begin{aligned}
 \overline{1} + \overline{30}  & \rightarrow \overline{31} \\
 \overline{3} + \overline{31}  & \rightarrow \overline{34} \\
 \overline{34}  & \rightarrow \overline{4} + \overline{30}
\end{aligned} \end{cases} \end{equation*}
They constitute a collectively-catalytic system. Then the reaction $\overline{1} + \overline{3} \rightarrow \overline{4}$ which is the overall reaction of the three can still be considered as low-barrier, and the formose reaction system is still self-replicating. The only question is that we have to wait for the complex molecule $\overline{30}$ to appear.

\section*{Discussion}

We have set up a general model for chemical reaction systems that properly accounts for energetics, kinetics and the conservation law. Although our model did not explicitly include catalysts, as other models did \cite{Hordijk17CTT, Hordijk04DAS, Steel00TEO}, catalysis and autocatalysis emerge in a number of systems.

We found three distinct types of self-driven system, i.e., systems which ``feed'' themselves. Both collectively-catalytic and self-replicating systems are vital in biology, while the third (non-sustaining) system appears less important. In terms of generating complexity, self-replicating system plays a more important role, since it is able to replicate innovations. In the self-replicating formose reaction \eqref{eq:formose}, after the first molecule $\overline{2}$ is produced by a high-barrier reaction, more $\overline{2}$s are easily replicated. In a biological setting, if this molecule spread to other places, it can trigger more self-replicating system. In contrast, in the collectively-catalytic citric acid cycle \eqref{eq:Krebs} for example, after the innovation (the first molecule $\overline{5}$), the second $\overline{5}$ will not appear until the responsible high-barrier reaction occurs once again.

By arbitrarily constructing alternative chemical universes, we found that lots of them contain self-driven systems, and the lower bounds on the number of chemical universes containing collectively-catalytic or self-replicating systems increase with more types of molecules. This result suggests that in a random chemical universe, it would not be too surprising to observe the emergence of self-replication, one of the central properties of life \cite{Szathmary06TOO}. Although it is not the first theory to propose that self-replication is relatively easy to emerge, as the RAF theory did \cite{Kauffman86ASO, Hordijk13ASF, Hordijk12TSO}, it is the first one requiring no catalyst.

We provided a general model explicitly showing that high thermodynamic free energy molecules can be produced exponentially from low free energy molecules, while specific mechanisms in specific real-world scenarios have been investigated before \cite{Amend13TEO, Branscomb13TAB}. The example system we showed is a metaphor of why high free energy ATP molecules are constantly produced in organisms \cite{Kun08CIO}. In addition, as our model takes energetics (corresponding to entropy) into account, it provides a more concrete way to discuss the issue---famously put forward by Schr\"odinger \cite{Schrodinger44WIL}---why life is able to spontaneously maintain a relatively low entropy level, although it cannot give the full answer: Answering the questions of how the molecules in living systems can be placed in an ordered structure (namely, a low entropy state) would require extending our model to include spatial effects.

Our model explicitly shows that complexity evolves from extreme simplicity stage by stage. It gives insights into three issues related to the origin of life. Firstly, the first RNA molecule is much more likely to be produced \textit{de novo} by this stage-by-stage procedure, rather than a magic event \cite{Robertson12TOO}. It provides a theoretical support to ``metabolism-first'' theories \cite{Robertson12TOO}, such as W\"achtersh\"auser's iron-sulfur world hypothesis \cite{Wachtershauser88BEA} and Szathm\'ary's theory \cite{Szathmary99Level, Szathmary95ACO, Maynard95TMT}. Secondly, before life, Earth should have gone through many stages in which different self-replicating systems existed and consequently Earth's compositions were different in each stage. The raw materials for life we should look for are those for the first self-replicating system, rather than those for the extant life \cite{Wachtershauser88BEA}. That is why, in the current theoretical framework, the raw materials for life (e.g., nucleotides) seem not to be available on the primordial Earth \cite{Robertson12TOO, Kim11SOC}. Thirdly, collectively-catalytic and self-replicating systems generate more types of new molecules, and in return, more types of molecules make more reactions feasible (in the form of catalysis and autocatalysis). This could explain why metabolic reactions in extant life always require sophisticated enzymes \cite{Sousa15ASI}, while no reaction is expected to involve catalysts in the very early stage of life \cite{Sievers94SOC, Rasmussen16GML, Levy03EGB}.

The model on its own provides a convenient platform to construct alternative chemical universes and investigate general properties of chemical reaction systems. It may provide a theoretical guideline for systematically searching for other chemical paths towards life (or at least self-replicating entities), as pursued in astrobiology \cite{Cavicchioli02EAT, Mix06TAP} and xenobiology \cite{Schmidt10XAN} for example. However, there are limitations. To be as simple as possible, our model assumes molecules with the same mass identical. As a result, in principle, every type of molecule can be produced from other molecules. But in reality it is not always the case, e.g., organometallic compounds can never be produced by a chemical reaction system involved with only carbohydrates. Nonetheless, the current model can be considered as a simple version of a more general model where conservation of different ``atoms'' is considered.





\bibliography{selfdriven}

\end{document}